\documentclass{appolb}
\usepackage{epsfig}

\newcommand{\beq}{\begin{equation}}
\newcommand{\eeq}{\end{equation}}
\newcommand{\bea}{\begin{eqnarray}}
\newcommand{\eea}{\end{eqnarray}}

\renewcommand{\l}{\lambda}
\newcommand{\La}{\Lambda}

\newcommand{\tr}{\mathrm{tr}\,}

\newcommand{\prt}{\partial}

\newcommand{\equ}{\!=\!}

\newcommand{\cN}{{\cal N}}

\newcommand{\cO}{{\cal O}}

\newcommand{\non}{\nonumber \\}

\newcommand{\eqref}[1]{(\ref{#1})}

\newcommand{\figref}[1]{Fig.~\ref{#1}}

\def\runhead{\ifPreprint \fi}

\begin{document}
\title{A Causal Alternative for $c=0$ Strings
\thanks{Presented at ``The 48th Cracow School of Theoretical Physics: Aspects of Duality'', 
June 13-22, 2008, Zakopane, Poland.
}%
}
\author{ J.\ Ambj\o rn$\,^{a,b}$, R.\ Loll$\,^{b}$,
Y.\ Watabiki$\,^{c}$, W.\ Westra$\,^{d}$ and \\ 
S.\ Zohren$\,^{e,f}$
\address{
$^a$~The Niels Bohr Institute, Copenhagen University\\
$^b$~Institute for Theoretical Physics, Utrecht University, \\
$^c$~Tokyo Institute of Technology, Dept. of Physics, High Energy Theory Group \\
$^d$Department of Physics, University of Iceland\\
$^e$Mathematical Institute, Leiden University\\
$^f$Blackett Laboratory, Imperial College\\
}
}
\maketitle
\begin{abstract}
We review a recently discovered continuum limit for the 
one-matrix model which describes ``causal''  two-dimensional quantum gravity. 
The behaviour of the quantum geometry in this limit is  
different from the quantum geometry of Euclidean two-dimensional
quantum gravity defined by taking the ``standard'' continuum 
limit of the one-matrix model. Geodesic distance and time scale
with canonical dimensions in this new limit, contrary to the situation
in Euclidean two-dimensional quantum gravity. 
Remarkably, whenever we compare, the known results of (generalized) 
causal dynamical triangulations are reproduced exactly by the 
one-matrix model. We complement 
previous results by giving a geometrical interpretation of the new 
model in terms of a generalization of the loop equation of Euclidean 
dynamical triangulations. In addition, we discuss the time evolution 
of the quantum geometry.
\end{abstract}

\section{Introduction}

Two dimensional quantum gravity is an interesting playground for 
quantum geometry. General ideas for  string theory and quantum gravity 
can be tested by exactly solvable models.
Particularly for string theory the $2d$ gravity point of view has 
been rather fruitful. 
Following the seminal work of Polyakov 
et al.~\cite{Polyakov:1981rd} 
powerful conformal field theory methods were developed. They allowed an exact solution 
for a select class of observables of 2d gravity 
(see e.g.~\cite{Fateev:2000ik}), including the coupling of simple matter 
models. 

Starting earlier \cite{Tutte1962a} 
and further developed in parallel with the continuum methods 
\cite{Ambjorn:1985az} 
is the method of dynamical triangulation (DT). 
Especially (generalized) enumeration of the DT 
random surfaces by matrix models proved fruitful, and random matrix models
became an important tool in the study of 2d Euclidean quantum gravity
coupled to certain conformal field theories.
Morever, the current understanding is that whenever the discrete and 
continuum methods can be compared the results coincide \cite{Martinec:2003ka}.

For a large class of observables the discrete methods have been proven 
to be more powerful. Correlators with surfaces of higher genus can 
be efficiently computed by matrix model techniques 
\cite{Ambjorn:1992gw} 
and fixed geodesic distance (propertime) correlation functions 
can be extracted by loop equations \cite{Watabiki:1993ym}, 
transfer matrices \cite{Kawai:1993cj} or through bijections 
with random trees \cite{Schaeffer1997}. 
So far these results have eluded the continuum methods.

In 1998 a different theory of 2d gravity was introduced 
called causal dynamical triangulations (CDT) \cite{Ambjorn:1998xu}. 
Using computer simulations the method has been successfully extended 
to 4d quantum gravity 
\cite{Ambjorn:2004pw}. 
The results are very promising and indicate that four dimensional gravity 
might be nonperturbatively renormalizable. The origin of the 
renormalizability could be a nontrivial fixed point scenario as 
described by Weinberg \cite{Weinberg:1980gg}. 

Although similar in spirit to Euclidean DT, the continuum limit of 2d 
CDT is significantly different. The main cause that puts 2d CDT in a 
different universality class from non-critical string theory is that in 
CDT the topology of spatial slices is fixed. This makes generic 
triangulations in CDT much better behaved, since unlike in Euclidean DT 
the spatial topology fluctuations cannot dominate the continuum limit. 
Consequently, the scaling of time in Euclidean DT is non-canonical and 
the Hausdorff dimension of the quantum geometry is $d_H\equ 4$. The 
quantum geometry of CDT on the other hand has a canonically scaling 
time variable and a Hausdorff dimension of $d_H\equ 2$.

Recent developments have shown that spatial topology fluctuations 
{\it can} be included while preserving the appealing features of CDT. 
The main idea is to assign a scaling coupling constant to the 
spatial topology change process \cite{Ambjorn:2007jm} and that 
this generalized CDT model can be described by a matrix model \cite{newmatrix}.

Before coming to the matrix model we rederive the disc function of pure CDT, 
i.e.\ without spatial topology change, by a simple geometrical loop equation.  
Such an equation is known to be remarkably powerful.  It allows one to 
compute time dependent correlators, where time is defined as the geodesic 
distance \cite{Watabiki:1993ym,Ambjorn:1999fp}. At the end of this letter 
we derive the differential equation for the time dependent propagator and 
show that unlike in Euclidean DT, but typical for CDT, the scaling of time 
is canonical.

After discussing pure CDT we add a term to the loop equation that 
introduces spatial topology change. Only upon adding this term we 
can relate the loop equation to the Schwinger-Dyson equation of a 
one-matrix model with a linear term in the action.  Remarkably, 
the linear term allows us to take a continuum limit that is very 
different from the well known limit of Euclidean DT but very similar 
to the continuum limit of CDT \cite{newmatrix}.  In fact, the continuum 
limit of the generalized loop equation reproduces the results of 
\cite{Ambjorn:2007jm,Ambjorn:2008ta}. 
Amazingly, the continuum limit of the matrix model can already be 
taken at the level of the matrix action, giving another matrix model that 
has a direct continuum interpretation \cite{Ambjorn:2008matrix}.




\section{Geometrical loop equations for 2D causal quantum gravity}

In this section we compute the generating function for 
a set $\cN$  of triangulations which are similar to 
the original set of causal triangulations and which lead 
to the same continuum physics: Let  $n$
denote the number of triangles and $l$ the number of 
links at the boundary (which has one marked link), and assume
the topology is that of the disk and denote the generating function
$\Phi(g,x)$:
\beq
\Phi(g,x)=\sum^{\infty}_{l,n=0} [\Phi(g,x) ]_{n,l} =\sum^{\infty}_{l,n=0} \cN (n,l) g^n x^l=\sum_{l=0}^\infty p_l(g) x^l.
\eeq
\begin{figure}[t]
\begin{center}
\includegraphics[width=5in]{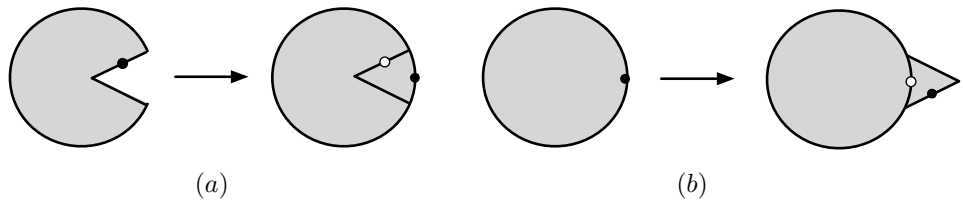}
\caption{Illustration of the two composition moves to add a triangle. The white dot on the right-hand-side shows the position of the mark before the triangle was added whereas the black dot shows the mark after the triangle was added.}
\label{fig0}
\end{center}
\end{figure}
The triangulations can be generated by recursively adding triangles. 
In our model there are two possible moves.
Firstly, one can glue two edges of the additional triangle to the 
triangulation, one to the marked edge and the other one next to it in 
the clockwise direction (\figref{fig0} (a)). Secondly, one can add a triangle by 
simply gluing one of its edges to the marked edge of the triangulation and assigning the new mark to the new edge further clockwise (\figref{fig0} (b)).  
Together, the two moves give the following generating equation for 
large $n$ and $l$ (see \figref{fig1}),
\beq \label{eq:le1}
[\Phi(g,x) ]_{n,l}=  \frac{g}{x} [\Phi(g,x) ]_{n,l} +  g x[\Phi(g,x)]_{n,l}.
\eeq
To keep the nice 
pictorial interpretation whilst making equation \eqref{eq:le1} exact, 
even for $n,l=1$ and $n,l=0$, one defines the following derivative operator, 
see e.g.\ \cite{Carroll:1995nj},
\beq
\prt_x\sum_{l=0}^\infty c_l x^l=\sum_{l=1}^\infty c_l x^{l-1},
\eeq
\beq
\prt_x \Phi = \frac{1}{x}\left(\Phi-1\right),~~~~~~~
\prt^2_x \Phi = \frac{1}{x^2}\left(\Phi - x p_1(g)- 1 \right).\nonumber
\eeq
The exact  generating equation can now be written as
\beq \label{eq:le1b}
\Phi(g,x)= 1  + g x + g x \prt^2_x \Phi(g,x) + g x^2 \prt_x \Phi(g,x),
\eeq
 where $x$ and $\prt_x$ have the clear graphical interpretation of 
adding and removing boundary edges.
\begin{figure}[t]
\begin{center}
\includegraphics[width=4in]{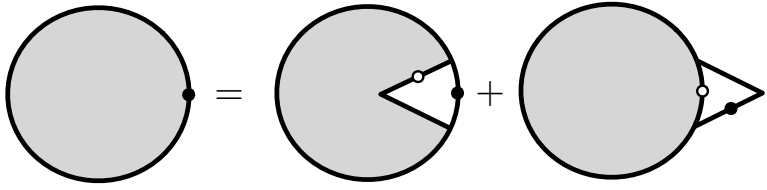}
\caption{Graphical representation of the loop equation \eqref{eq:le1} and \eqref{eq:le1b}.}
\label{fig1}
\end{center}
\end{figure}
Equation \eqref{eq:le1b} is a simple linear equation and the solution 
is given by
\beq \label{eq:phires1}
\Phi(g,x)= g \left(\frac{1 - (1/g  - p_1(g)) ~ x }{g x^{2} - x + g  }\right),
\eeq
where $p_1(g)$ can be determined by demanding that the singularity 
structure of \eqref{eq:phires1} does not change discontinuously near $g=0$. 
The poles of \eqref{eq:phires1}  are located at
\beq
x_{\pm} = \frac{1 \pm \sqrt{1-4 g^2}}{2 g},~~~~~~~1/g - x_- = x_+.
\eeq
Since the expansion of $p_1(g)$ needs to be a power series we have 
that  $p_1=x_-$, hence the disc function is given by 
\beq
\Phi(g,x)= \frac{1}{1-p_1(g)\, x} ,~~~~~~~p_l(g)= p_1(g)^l.
\eeq

\section{The continuum limit}

Using the same scaling relations as in the transfer matrix formalism of CDT,
\beq \label{eq:pureCDTscaling}
g=\frac{1}{2}e^{-a^2 \La /2},\quad x=e^{-a X},
\eeq
we reproduce the continuum disc amplitude of causal quantum 
gravity \cite{Ambjorn:1998xu}
\beq
W_{\La}(X) = \frac{1}{X+\sqrt{\La}},~~~~~~~W_{\La}(L)=e^{-\sqrt{\La}L}.
\eeq

\section{A matrix model for generalized 2D causal quantum gravity}

\begin{figure}[t]
\begin{center}
\includegraphics[width=2in]{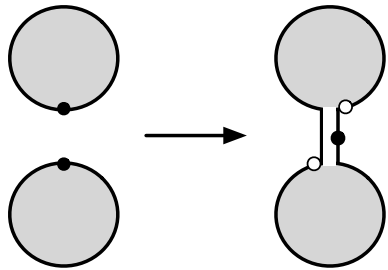}
\caption{Illustration of the composition moves to add a double link.}
\label{fig3}
\end{center}
\end{figure}

\begin{figure}[t]
\begin{center}
\includegraphics[width=5in]{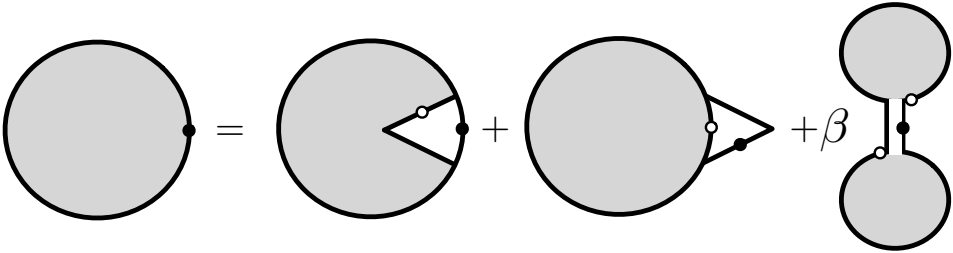}
\caption{Graphical representation of the loop equation \eqref{eq:le2}.}
\label{fig2}
\end{center}
\end{figure}

To include spatial topology change  we introduce a quadratic term in the 
loop equation \eqref{eq:le1b} (see \figref{fig3})
\bea
\Phi_\beta(g,x)= 1  + g x + gx \prt^2_x \Phi_\beta(g,x) +
       g x^2 \prt_x \Phi_\beta(g,x) + \beta x^2 \Phi_\beta(g,x)^2, \label{eq:le2}
\eea
where $\beta$ is a coupling constant that determines the rate of the 
spatial topology fluctuations (see \figref{fig2}).
To conform with matrix model conventions it is useful to introduce 
the following notation

\beq
w_\beta(g,z)= \frac{\Phi_\beta(g,x=1/z)}{z}.
\eeq
With these conventions the loop equation is given by 
\beq \label{eq:genloopmat}
\beta w_\beta(g,z)^2  -v' (z)w_\beta(g,z) +  q_\beta(g,z) = 0,
\eeq
where
\beq \label{eq:vdef}
v(z) = -g z + \frac{1}{2} z^2 - \frac{1}{3} g z^3,~~v'(z)= -g  +  z - g z^2,
\eeq
and
\beq \label{eq:qdef}
q_\beta(g,z) = 1  - g (p_1(g,\beta) + z).
\eeq
Written in the form of  \eqref{eq:genloopmat} it is easily seen that 
the loop equation corresponds
to the Schwinger-Dyson equation of a simple one-matrix model \cite{newmatrix},
\beq \label{eq:discmat}
Z_{disc.}=\!\int \!Dm \exp\left(\! -\frac{N}{\beta}\! \left[ \tr v(m) \right] \right),
\eeq 
where $m$ is a Hermitian $N \times N$-matrix and the functional form 
of the potential $v(m)$ is given by \eqref{eq:vdef}.

The solution of the loop equation \eqref{eq:genloopmat}  is of the 
following well-known form  
\beq
w_\beta(g,z) = \frac{1}{2 \beta} \left( v'(z) - \sqrt{v'(z)^2-4 \beta q_\beta(g,z)}\right).
\eeq
At this stage the solution of the disc function is still implicit since 
it depends on $p_1(g,\beta)$ through \eqref{eq:qdef}.
Demanding the solution to have only one cut in the complex $z$ plane 
gives the explicit result 
\bea \label{eq:wdiscrgen}
w_\beta(g,z) = \frac{1}{2 \beta} \left( -g  +  z - g z^2 +
(g z-c)\sqrt{ (z - c_+)(z - c_-)}\right),
\eea
where $c$ is the solution of a third order polynomial,
\beq \label{eq:polyc}
2 c^3 - 3 c^2 +\left(2 g^2 + 1\right) c =  g^2 (1-2 \beta ),
\eeq
and
\beq
c_{\pm}=\frac{1-c\pm\sqrt{2} \sqrt{(1-c) c-g^2}}{g}.
\eeq

\section{The ``causal'' continuum limit}

In the well known continuum limit of the one-matrix model with 
polynomial potential and generic coupling constants the critical 
value for the boundary cosmological constant $z$ coincides with 
the critical value of $c_+$ only.  As a result, this standard limit 
has the peculiar feature that it leaves a non-scaling term as a memory 
of the discrete theory, since 
\beq
w^{Euc.}_{\l}(x) =  w_{ns}(x) +a^{\frac{3}{2}} \:W^{Euc.}_{\La}(X) + \cO(a^2),
\eeq
where $w_{ns}(x)$ is a non-scaling part and
\beq
W^{Euc.}_{\La}(X) =  \left(X-\frac{\La}{2} \right) \sqrt{X+\sqrt{ \La}}
\eeq 
is the continuum disc function.
Observe that the continuum disc function is even subleading in the lattice 
cutoff $a$.

In our specific model the matrix potential is such that the critical points 
of $c_+$ and $c_-$ coincide.
This leads us naturally to the universality class of two-dimensional CDT 
implying the same scaling relations as before \eqref{eq:pureCDTscaling}, 
provided one also scales the coupling constant $\beta$ \cite{Ambjorn:2007jm}:
\bea \label{eq:genCDTscaling}
\beta& =&\frac{1}{2}g_s a^3,~~~~~~~\,c =\frac{1}{2}e^{ a C}.
\eea
Contrary to the standard limit of the one matrix model our 
``causal'' continuum limit is free from leading nonscaling contributions,
\beq
w_{\beta}(g,x) =  \frac{1}{a}  \:W_{\La,g_s}(X) + \cO(a^0),
\eeq
where $W_{\La,g_s}(X) $ is the continuum disc function previously 
derived with other methods 
\cite{Ambjorn:2007jm,Ambjorn:2008ta}, 
\beq \label{eq:contdiscgenral}
W_{\La,g_s}(X) = \frac{1}{2 g_s} 
\left( -(X^2-\La)+(X-C)\sqrt{(X + C)^2-2g_s/C}\right),
\eeq
where $C$ is the solution to a third order polynomial equation,
\beq
C^3 - \La C + g_s=0.
\eeq
Observe that this equation is precisely the continuum limit of 
equation \eqref{eq:polyc}. Furthermore it is interesting to note 
that the structure of the discrete \eqref{eq:wdiscrgen} and 
continuum \eqref{eq:contdiscgenral} disc functions is very similar. 
This is not a coincidence since, as has been noticed recently, 
the continuum  results can also be described by a matrix model 
\cite{Ambjorn:2008matrix},
\beq \label{eq:contmat}
Z_{cont.}=\!\int \!DM \exp\left(\! -\frac{N}{g_s}\! 
\left[ \tr V(M) \right] \right),
\eeq
with the following potential
\beq
V(M) = \La M \!-\! \frac{1}{3}  M^3.
\eeq
In fact, it can be shown that the continuum matrix model 
\eqref{eq:contmat} can be understood as the continuum limit of 
the matrix model \eqref{eq:discmat} with a standard combinatorial 
interpretation \cite{newmatrix}. 
While our continuum limit \eqref{eq:pureCDTscaling} and 
\eqref{eq:genCDTscaling} is non-standard from the matrix model 
point of view it is very natural from a CDT perspective 
\cite{Ambjorn:2007jm,Ambjorn:2008ta}.

\section{Time evolution}

To see that the non-critical string theory limit of the matrix model 
is really very different from our ``causal'' continuum limit, 
we briefly discuss the so-called fixed time (geodesic distance) 
two-loop amplitude $G_\beta(l_1,l_2;g;t)$. This amplitude is defined 
as the sum over all triangulations with initial boundary of 
length $l_1$ and final boundary of length $l_2$ at fixed geodesic 
distance $t$. With this in mind, it is natural to interpret the loop 
equation as a time dependent process 
\cite{Watabiki:1993ym,Ambjorn:1999fp,Arnsdorf:2001wh}, 
where each addition or subtraction of a triangle 
is a ``$1/l_1$-th'' part of a time step. For large $l_1$ we have 
\bea
&&\frac{1}{l_1}\frac{\partial}{\partial t}G_\beta(l_1,l_2;g;t) =g  G_\beta(l_1-1,l_2;g;t)-G_\beta(l_1,l_2;g;t) +\non 
&&\,\, + \,\, gG_\beta(l_1+1,l_2;g;t)+ 2  \beta\sum_{l=0}^\infty p_l(g,\beta) G_\beta(l_1-l-2,l_2;g;t).
\eea
After a ``discrete Laplace transformation'' this equation becomes
\beq    \label{eq:timeloop}
\frac{\partial}{\partial t} G_\beta(z,w;g;t)=
\frac{\partial}{\partial z}\left[ (-g+z-gz^2 -2 \beta w_\beta(g,z)) 
G_\beta(z,w;g;t)\right].
\eeq
The crucial difference between this equation and similar equations  
in non-critical string theory is the scaling of time in their 
continuum limits. Unlike in non-critical string theory the continuum 
limit  of \eqref{eq:timeloop} involves a canonically scaling time 
parameter $T\!\sim\!a\,t$, yielding
\beq\label{3.5newxxx}
 \frac{\prt}{\prt T} G_{\La,g_s}(X,Y;T) \!=\! 
- \frac{\prt}{\prt X} \Big[\left( X^2\! -\!\La \!+ \! 2 g_s W_{\La,g_s}(X) \right)\! G_{\La,g_s}(X,Y;T)\Big].
\eeq
This is precisely the result of the propagator derived from generalized 
CDT \cite{Ambjorn:2007jm,Ambjorn:2008ta}. Already since the inception of 
CDT \cite{Ambjorn:1998xu} it has been known that the scaling of the 
geodesic distance is intimately related to the Hausdorff dimension of 
the quantum geometry. One can argue that the ``causal'' continuum limit 
of the one-matrix model is better behaved since its Hausdorff dimension 
is $d_H\equ 2$ instead of  $d_H\equ 4$ in non-critical string 
theory \cite{Ambjorn:1998xu}.



\section{Conclusions}

We have described a recently found ``causal'' continuum limit 
for the one-matrix model \cite{newmatrix}. 
With the here described combinatorial interpretation this limit is naturally defined when inserting a linear term in the action with a specially chosen coefficient.
We have shown here that this coefficient can naturally be 
interpreted as an additional way to add a triangle in the loop equations. 
The associated ``causal'' continuum limit is very different from the ``old'' 
double scaling limit and exactly reproduces the known results of causal 
dynamical triangulations \cite{Ambjorn:1998xu,Ambjorn:2007jm}. 
An intriguing aspect of this new continuum limit is that the continuum 
results are also described by a matrix model.  
This matrix model has of course both an expansion in the coupling
constants $g_s$ and $\La$, as well as a large $N$ 
expansion in powers of $1/N^2$, which reorganizes the power expansions 
in $g_s$ and $\La$ in convergent ``subseries''. By comparing 
with the generalized causal dynamical triangulation model the 
powers of $N^{-2h+2}$ in the large $N$ expansion can be identified with
the continuum causal dynamical surfaces of genus $h$ 
\cite{Ambjorn:2008matrix,newmatrix}. In this sense our ``causal'' 
continuum limit leads to a picture that is much closer in spirit to the 
original idea by 't Hooft \cite{Hooft:1973jz} for QCD. 

As a complement to previous results we also derived the differential 
equation for the fixed geodesic distance two-loop amplitude. 
The observed canonical scaling of time is directly related to the 
fact that the quantum geometry obtained from our new ``causal'' 
continuum limit has Hausdorff dimension $d_H\equ 2$ instead of $d_H\equ 4$ 
for the ``old'' continuum limit of the one-matrix model.

Until now, only the transfer matrix formalism has been available for 
analytical computations in causal dynamical triangulations. 
Here we extended the available tools and presented new, 
more powerful, matrix model and loop equation methods.
Importantly, these new methods allow one to analytically study 
simple matter models coupled to two-dimensional causal quantum gravity. 
Several of these models are currently under investigation.\\

{\it Acknowledgments.---}
JA, RL, WW and SZ acknowledge the  support by
ENRAGE (European Network on
Random Geometry), a Marie Curie Research Training Network in the
European Community's Sixth Framework Programme, network contract
MRTN-CT-2004-005616. RL acknowledges
support by the Netherlands
Organisation for Scientific Research (NWO) under their VICI
program.

\end{document}